%% file: DIF_main.tex
\documentclass{edm_template}

\usepackage{graphicx}

\usepackage{tabularx}
\usepackage{booktabs}
\usepackage{array}
\usepackage{threeparttable}

\usepackage{accents}
\usepackage{relsize}
\usepackage{dblfloatfix}
\usepackage{array}
\usepackage{makecell} 
\usepackage{xurl}            
\usepackage[hidelinks]{hyperref}  
\usepackage{xcolor}

\newcolumntype{P}[1]{>{\raggedright\arraybackslash}p{#1}}
\newcolumntype{Y}{>{\raggedright\arraybackslash}X}

\makeatletter
\renewcommand{\theparagraph}{\thesubsection.\arabic{paragraph}}
\@addtoreset{paragraph}{subsection}
\makeatother
\makeatletter
\def\@copyrightspace{}
\makeatother

\title{Assessment Design in the AI Era: A Method for Identifying Items Functioning Differentially for Humans and Chatbots}

\begin{document}

\numberofauthors{6}
\author{
\alignauthor Licol Zeinfeld\textsuperscript{*}\\[0.55em]
Ron Blonder\\
\alignauthor Alona Strugatski\textsuperscript{*}\\[0.55em]
Shelley Rap\\
\alignauthor Ziva Bar-Dov\\[0.55em]
Giora Alexandron\\
\and
\alignauthor \vspace{0.9em}\centering
{\large \makebox[\linewidth][c]{Weizmann Institute of Science}}\\[0.40em]
{\small \makebox[\linewidth][c]{\textsuperscript{*}Equal contribution}}
}

\maketitle

\begin{abstract}
The rapid adoption of large language models (LLMs) in education raises profound challenges for assessment design. To adapt assessments to the presence of LLM-based tools, it is crucial to characterize the strengths and weaknesses of LLMs in a generalizable, valid and reliable manner. However, current LLM evaluations often rely on descriptive statistics derived from benchmarks, and little research applies theory-grounded measurement methods to characterize LLM capabilities relative to human learners in ways that directly support assessment design. Here, by combining educational data mining and psychometric theory, we introduce a statistically principled approach for identifying items on which humans and LLMs show systematic response differences, pinpointing where assessments may be most vulnerable to AI misuse, and which task dimensions make problems particularly easy or difficult for generative AI. The method is based on Differential Item Functioning (DIF) analysis -- traditionally used to detect bias across demographic groups -- together with negative control analysis and item-total correlation discrimination analysis. It is evaluated on responses from human learners and six leading chatbots (ChatGPT-4o \& 5.2, Gemini 1.5 \& 3 Pro, Claude 3.5 \& 4.5 Sonnet) to two instruments: a high school chemistry diagnostic test and a university entrance exam. Subject-matter experts then analyzed DIF-flagged items to characterize task dimensions associated with chatbot over- or under-performance. Results show that DIF-informed analytics provide a robust framework for understanding where LLM and human capabilities diverge, and highlight their value for improving the design of valid, reliable, and fair assessment in the AI era.
\end{abstract}

\keywords{Assessment, Generative AI, Differential Item Functioning}

\section{Introduction}
The rapid adoption of generative AI (GenAI\footnote{In this paper, we use GenAI to refer to generative AI in the broad sense; LLMs to refer specifically to language-oriented models; and chatbots to refer to conversational agents powered by LLMs.}) tools in education has created both opportunities and risks. While these systems, particularly chatbots such as ChatGPT, can provide personalized explanations, feedback, and support for learners, their growing use also poses a profound threat to the validity of educational assessments \cite{The-AI-Cheating-Crisis, Cotton03032024}. By making it exceptionally easy for students to generate fluent and convincing answers, GenAI lowers the barriers to academic dishonesty in ways that traditional safeguards struggle to address \cite{The-AI-Cheating-Crisis, Cotton03032024, khalil2023will}. This risk is amplified in unproctored or remote environments, where monitoring is limited and misuse is harder to detect \cite{yan2024practical}. Importantly, the concern is no longer theoretical: recent studies document that students are already using GenAI for assignments and exams, directly undermining the validity and reliability of assessment results and raising urgent challenges for educational practice and policy \cite{Prothero24, susnjak2024chatgpt}. These changes point to a key observation: to ensure resistance to GenAI-related misconduct and to maintain validity and reliability in assessment contexts characterized by potentially heavy LLM use, it is necessary to understand where LLMs diverge from human learners and to characterize their capabilities and limitations across assessment tasks.

Current evaluations of LLM performance on assessment tasks are largely shaped by technical benchmarks (e.g., \cite{borges2024could,wang2023scibench,lu2022learn}). These benchmarks provide descriptive statistics that locate LLMs on various scales relative to human learners and provide evidence that LLMs are affected differently from humans by task features such as visual elements or sequential reasoning steps \cite{wang2024examining,yacobson2025benchmarking}. Yet, little research has examined more generalizable ways of measuring LLM capabilities in relation to human learners. Assessment, by definition, is a proxy that seeks to generalize from sparse samples of student performance (e.g., across time or content). For human learners, the design and interpretation of such proxies have been refined through decades of assessment research and implementation. For GenAI agents, however, it remains far less clear what inferences assessments support.

Mature frameworks from educational measurement -- such as, but not limited to, Item Response Theory (IRT) \cite{DeAyalaIRT} -- may offer principled ways to assess the capabilities of GenAI. Recent studies \cite{sorenson2024identifying, Strugatski_25} have shown that psychometric modeling can be extended beyond its traditional applications to capture systematic differences between human learners and chatbots, mainly for identifying GenAI. Within educational data mining, a central paradigm is that item-level analysis can reveal underlying cognitive processes, providing a foundation for more valid assessment and adaptive learning design \cite{barnes2005q, koedinger2012automated}. Combining these approaches, our work moves the center of attention from \textit{examinees} to \textit{items}, and its goal is identifying items on which chatbots and humans exhibit differential behavior. Identifying such items serves two important purposes. First, to better understand the dimensions that make tasks easy or difficult for LLMs relative to human learners. Second, to apply these insights to the design of assessments that are better adapted -- in terms of validity, reliability, and fairness -- to scenarios in which students may work (legitimately or not) with GenAI.

Within psychometrics, there is a collection of methods known as Differential Item Functioning (DIF). DIF refers to situations in which an assessment item functions differently for subgroups of learners distinguished by a characteristic unrelated to the construct being measured (typically a demographic characteristic). In DIF terminology, the baseline group is referred to as the \textit{reference} group, while the group of interest is referred to as the \textit{focal} group. DIF methods go beyond simply comparing overall performance between the reference and focal groups by controlling for student ability, \texttt{thereby distinguishing true item bias from general group-level performance differences}.

DIF analysis is widely used in assessment design to flag biased or poorly constructed items, as such items compromise the validity and fairness of the assessment \cite{Martinkov2017DIF}. For example, a mathematics item requiring advanced reading comprehension may disadvantage learners with equivalent math ability but weaker language skills (e.g., second language learners). Methods to detect DIF include non-parametric methods, such as Mantel--Haenszel \cite{mantel1959statistical}, or parametric ones, such as those based on IRT or logistic regression \cite{magis2010general}.

The rationale of the current work is to examine whether DIF methods can be useful to identify items on which human learners and chatbots differ. The observation is that chatbots can be referred to as the `focal' group, while humans are the `reference' (baseline) group. This is formulated through the following research questions (RQs):

\noindent\textit{\textbf{RQ1:} Can DIF techniques identify items that function differently for human learners and chatbots?}

\noindent\textit{\textbf{RQ2:} What are the key characteristics that subject-matter experts identify in items on which chatbots exhibit differential performance compared to human learners?}

To study these questions, we developed, in a stepwise manner, a method that combines DIF analysis with additional item-level criteria and identifies items on which chatbots and human learners show differential behavior. This method was progressively refined and evaluated on human learners and GenAI responses to two assessment instruments taken from two very different contexts: a high school chemistry test administered as a formative assessment, and the quantitative section of a high-stakes psychometric entrance exam for higher education. The GenAI responses were generated using six chatbots (see the Methodology section for details). Following that, subject-matter experts conducted a qualitative analysis of the chemistry DIF items to identify key dimensions that may lead to the differential behavior.

The key contribution of this paper is proposing a theory-inspired and statistically robust method to identify items that exhibit differential behavior for chatbots relative to students, and demonstrating its effectiveness in providing item analytics to those who wish to incorporate GenAI-related considerations into assessment design.

\input{sections/methodology.tex}

\input{sections/experiments_and_results.tex}

\section{Discussion}
The LR-based DIF approach developed and piloted in this research provides a method and conceptual framework for identifying assessment items that show differential behavior between humans and chatbots. In two assessment contexts and on leading chatbots, we demonstrated that the method provides reliable and stable results. A key observation that DIF, and this research, makes, is that differential behavior is a more nuanced property than merely looking at overall performance gaps between groups, since the DIF analysis controls for ability level, and can identify items as positive/negative DIF (chatbots over/under-perform learners of the same skill) even if the difference between the mean group performance is in the opposite direction. Our LR-DIF based method offers a modeling that also enables the detection of non-uniform DIF patterns in which either the magnitude or direction of group differences vary across ability levels.
Our experiments with both the MH- and LR-DIF highlighted that the MH-DIF, typically the default choice due to its simplicity, produces considerable noise and a high false positive rate. The LR-DIF excelled in this aspect as well. Its downside is its relative complexity and the fact that it typically requires larger sample sizes for stable estimates.

Using the DIF-method to deliver content analytics to the subject-matter experts (RQ2) drove an analysis that yielded interesting insights. As reported in Subsection~\ref{sec:sme}, analyzing the POS DIF items, we found that the chatbots managed to circumvent distractors aimed at surfacing alternative conceptions (sometimes referred to as `misconceptions') commonly held by students. Given that LLMs can be cognitively biased due to their training data \cite{echterhoff-etal-2024-cognitive}, which likely included student data also representing wrong or incomplete knowledge \cite{yan2024practical}, the fact that the chatbots dodged these designated `traps' is interesting, and is in disagreement with previous work that found moderate alignment between LLMs and student misconceptions \cite{liu2025llms}. On the contrary, experts' analysis of the NEG DIF items revealed that the chatbots underperformed on items requiring visual interpretation and connecting visual and textual information, items that their wording required understanding linguistic nuances, and those requiring performing multi-step problem solving procedures. These findings reinforce previous work about GenAI problem solving in STEM domains \cite{wang2024examining,yacobson2025benchmarking}.

More generally, this research demonstrates a fruitful application of assessment and measurement theory to educational data mining, with the purpose of establishing robust approaches for understanding GenAI capabilities, and for developing assessment in the GenAI era.

\noindent\textbf{\textit{Limitations.}}
This research is the first, to our knowledge, to apply DIF for analyzing chatbot assessment data and, as such, naturally has several limitations. In terms of internal validity, the chatbots' bi-modal behavior may impact the DIF stratification step, potentially reducing the effective size of the data and thus the method's robustness. The main limitation to external validity is that the results are based on a small number of instruments and specific GenAI tools. 

\noindent\textbf{\textit{Contribution and future work}.}
The main contribution of this work is the introduction of a theory-driven and statistically sound method for detecting items that exhibit differential behavior between chatbots and human learners, and the presentation of evidence of its ability to yield meaningful analytics for subject-matter experts who seek to integrate GenAI considerations into assessment design. In future work, we plan to build on this foundation for (1) studying the task dimensions of Human--GenAI DIF items, to better understand the capabilities of this technology, (2) incorporating Human--GenAI DIF analysis into assessment design, and (3) Use DIF analysis to evaluate GenAI progress.

\vspace{-0.06em}
\noindent\textbf{Code availability.}
The full code and sample data to reproduce the analyses in the paper can be found in the GitHub repository: (\href{https://anonymous.4open.science/r/Assessment_Design_in_the_AI_Era-4A08/}{\textcolor{blue}{\underline{link}}})

\section{Acknowledgments}
This work was supported by the Knell Family Institute for Artificial Intelligence, Israel. The authors thank the National Institute for Testing and Evaluation for providing access to psychometric exam data.
\bibliographystyle{abbrv}
\bibliography{refs}

\end{document}

%% file: sections/methodology.tex
\section{Psychometric Preliminaries}\label{sec:pre}

\subsection{Item–Total Correlation and Rest Score}
Item–total correlation (ITC) is defined as the correlation between the score on a single item and the \textit{rest score} for that item, which is the aggregated performance across all the \textit{other} items in the test (also named `corrected ITC'). It assesses the consistency of an item with the rest of the test, providing a measure of \textit{item discrimination} -- how well the item distinguishes between examinees with high versus low overall performance on the test. ITC analysis is used in test design to improve validity and reliability. 
Following common practice, ITC values of 0.20 or higher are considered acceptable \cite{shete2015item}; we adopt this threshold in the present study.

\subsection{Differential Item Functioning (DIF)}\label{sec:dif}
DIF methods test whether an item functions differently across groups of respondents once respondents’ \emph{abilities} have been controlled for. Here, `ability' refers to a respondent's overall proficiency on the instrument and is measured by the per-item rest score, i.e., the sum of the respondent's correct responses across all items excluding the target item. In DIF terminology, the \emph{reference group} is the baseline for comparison (here, humans), and the \emph{focal group} is the group tested for differences (here, chatbots). For analysis, we denote each response as $X_{ij}\in\{0,1\}$ for respondent $i$ on item $j$, where 1 indicates a correct answer and 0 an incorrect one. Conceptually, an item exhibits DIF if two respondents with the same ability, but from different groups, have systematically different probabilities of answering it correctly. Throughout, we adopt a consistent direction convention: \textbf{POS}/\textbf{NEG} (positive/negative) indicates a chatbot advantage/disadvantage (higher/lower conditional probability of answering correctly than humans at the same ability level).

\paragraph{Mantel--Haenszel DIF (MH-DIF)}\label{sec:dif-MH}
MH-DIF compares focal and reference groups within discrete strata of the ability proxy and aggregates these comparisons into a common odds ratio ($\alpha_{\text{MH}}$). Odds ratios less than one correspond to POS (chatbot advantage), while odds ratios greater than one correspond to NEG (chatbot disadvantage). Following ETS guidelines \cite{zwick2012rr1208}, effect sizes are categorized by the magnitude of $|\log(\alpha_{\text{MH}})|$:  
Category~A (\textit{negligible}) if $|\log(\alpha_{\text{MH}})| < \log(1.5)$,  
Category~B (\textit{moderate}) if $\log(1.5) \leq |\log(\alpha_{\text{MH}})| < \log(2)$,  
and Category~C (\textit{large}) if $|\log(\alpha_{\text{MH}})| \geq \log(2)$.  
  MH-DIF is widely used for detecting \emph{uniform DIF}, or differences that are stable across the ability range, and is particularly robust under stratification with limited sample sizes. For more details on MH-DIF, see \cite{rogers1993comparison}.

\paragraph{Logistic Regression DIF (LR-DIF)}\label{sec:dif-LR}
LR-DIF provides a more flexible framework that models the probability of a correct response as a function of ability, group membership, and their interaction:
\[
\text{logit}\,P(X_{ij}=1) = \beta_0 + \beta_1 A_{i,-j} + \beta_2 G_i + \beta_3 (A_{i,-j}\times G_i),
\]
where $G_i=0$ for humans and $G_i=1$ for chatbots, and $A_{i,-j}$ is the per-item rest score used as an ability proxy. Nested model comparisons allow us to test separately for \emph{uniform DIF} (via the main group effect $\beta_2$) and \emph{non-uniform DIF} (via the interaction term $\beta_3$). In cases where the interaction implies different directions across the ability range, we label the effect as \emph{Sign-change DIF}. From the fitted logistic function we compute likelihood-ratio $p$-values ($m_0$ vs.\ $m_1$, $m_1$ vs.\ $m_2$), McFadden’s $\Delta R^2 = R^2_{m_2}-R^2_{m_0}$ with $R^2_{m}=1-\ell(m)/\ell(m_\text{null})$, and at-median odds ratios, which indicate how much more or less likely chatbots are to answer correctly relative to humans at the median ability level after trimming. We adopt $\Delta R^2 \geq 0.035$ as the threshold for a meaningful effect size \cite{JodoinGierl2001}, and use these measures together to detect and characterize DIF effects.

\section{Methodology}\label{sec:method}
This section describes how the psychometric measures above were applied to design a statistical method that identifies items that function differently for humans and chatbots (RQ1), and also the process through which DIF items were subjected to qualitative analysis by subject matter experts to characterize human–chatbot DIF behavior (RQ2).

\begin{figure*}[!t]
  \centering
  \includegraphics[width=\textwidth]{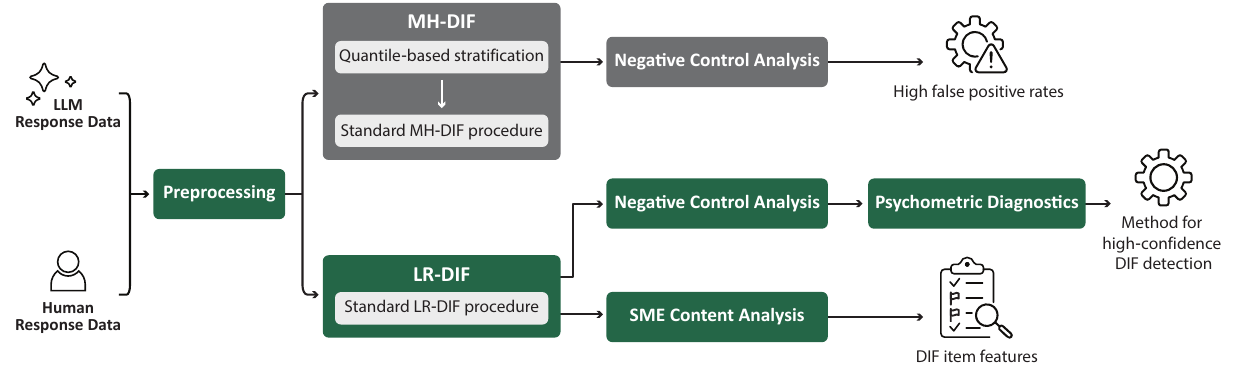} 
  \caption{Implemented methodological framework for human--LLM DIF analysis.}
  \label{fig:pipeline}
\end{figure*}

\subsection{Method Design}

\noindent\textit{\textbf{Task Definition and Modeling.}} Our task is to develop a statistically principled method that detects and characterizes multiple-choice items exhibiting systematic response differences between human learners and LLM-based chatbots. Adopting the DIF terminology, we refer to humans as the reference group and chatbots as the focal group.\newline
\noindent\textit{\textbf{Design Process.}}
 To develop a method for identifying true item-level differences between humans and chatbots, we followed a stepwise process, which is outlined in Fig.~\ref{fig:pipeline}. Below, we briefly describe the  key steps (the implementation  and results are detailed in Section~\ref{sec:results}): \newline
\noindent\textbf{Step 1 -- DIF Analysis:} Given response data from humans and chatbots (see data description below), we applied the two DIF methods defined above (MH- and LR-DIF). These methods allow us to identify items where humans and chatbots with the same ability level differ in the probability of providing a correct response (here, \emph{ability} refers to respondents' overall proficiency on the instrument). We interpret the items flagged in this step as candidate DIF items.\newline
\noindent\textbf{Step 2 -- Negative Control Analysis:} To control for false positives and validate the overall stability of the DIF analysis, we performed a negative control analysis (Placebo Test \cite{eggers2024placebo}). For that, Step 1 was reapplied to 50 null datasets per instrument, where the chatbot group was replaced with random samples of human responses (drawn without replacement). On each null dataset, both MH- and LR-DIF were rerun, and we recorded which items were flagged as DIF to compute their false-positive rates.\newline
\noindent\textbf{Step 3 -- Applying Psychometric Diagnostics:}
 Next, we analyzed the items flagged by the placebo test in Step 2 using psychometric validation measures, with the purpose of identifying whether these items have certain psychometric characteristics (e.g., low ITC) that can explain their instability and be used to filter them upfront. 


\subsection{Instruments and Data}\label{sec:data} 
We used two assessment instruments from two very different contexts. The first was a chemistry diagnostic test administered as a formative assessment activity in preparation for the high school matriculation exam. It consisted of 22 multiple-choice items answered by 931 students. The second was the quantitative section of a psychometric entrance exam used for higher-education admissions, containing 40 items with over 4,800 respondents. Both instruments featured multi-modal items with figures, images, and formulas. 

To generate chatbot data, we collected responses from six LLMs representing three distinct model families. For each family, we included both a previous version and the most recent release available at the time of study, specifically: OpenAI’s GPT-4o and GPT-5.2; Google’s Gemini 1.5 Pro and Gemini 3 Pro; and Anthropic’s Claude 3.5 Sonnet and Claude 4.5. Each instrument was uploaded with a prompt requesting only the final answers. The models typically returned numbered answer lists, which we exported to CSV. This process was repeated 20 times per model, simulating 20 ``artificial students'' for each chatbot. Runs were conducted in separate sessions to prevent memory effects, and variation across responses was introduced by using a non-zero default temperature setting. Since both assessments included multimodal content, it was important that all models supported visual inputs. When a model skipped an item or did not provide a valid option, the response was counted as incorrect (similar to how student responses were treated). The combined human and chatbot responses were then balanced by down-sampling the more abundant data source to construct a dataset at a 1:10 ratio.

\begin{figure*}[!t]
  \centering
  \includegraphics[width=0.75\textwidth]{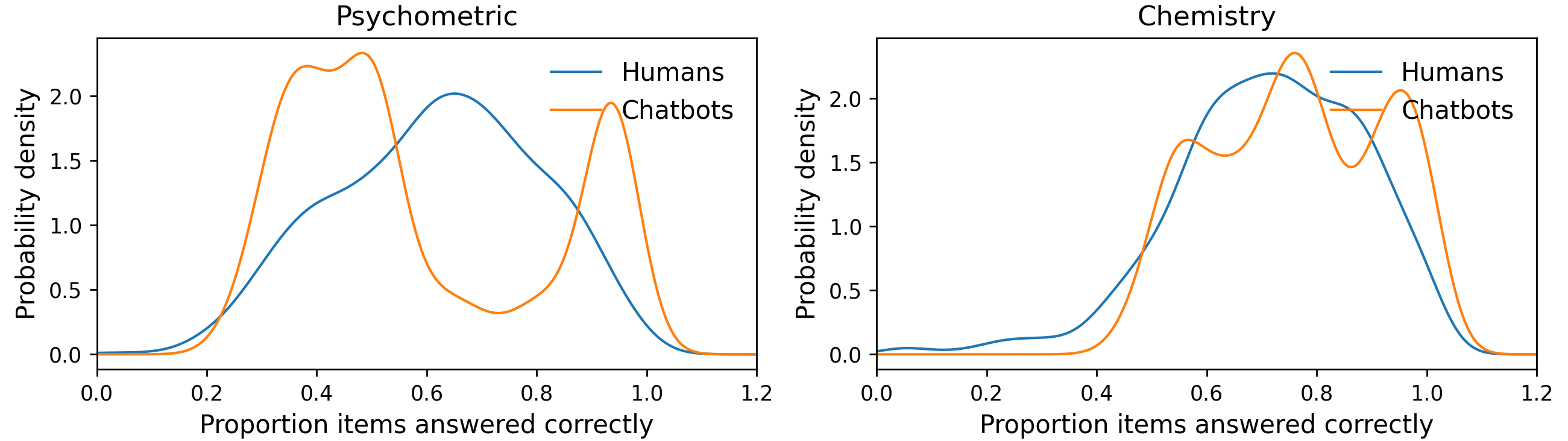}
  \caption{Human--Chatbot ability distribution.}
  \label{fig:ability-distribution}
\end{figure*}

The ability distributions of the chatbots and human learners on both instruments are presented in Fig.~\ref{fig:ability-distribution}. The bi-modal distribution of the chatbots' ability originates from the division between previous and recent chatbot versions (to examine the robustness of the method to model generations, we validated the method reported below also on a subset containing only the previous versions). Table~\ref{tab:avg_score} shows the fraction correct of each group on the chemistry items (which were further analyzed by the subject-matter experts). As shown, although the overall success rate of both groups is similar, the chatbots have higher success rates on most items. These sophisticated relations can be handled by the MH- and LR-DIF, as we later demonstrate.



\begin{table*}[!t]
\centering
\caption{Fraction correct per item for students and chatbots (chemistry).}
\label{tab:avg_score}

\scriptsize
\setlength{\tabcolsep}{1.3pt}      
\renewcommand{\arraystretch}{1.10}

\begin{tabular*}{0.985\textwidth}{@{\extracolsep{\fill}}l*{22}{c}}
\toprule
 & Q1 & Q2 & Q3 & Q4 & Q5 & Q6 & Q7 & Q8 & Q9 & Q10 & Q11 & Q12 & Q13 & Q14 & Q15 & Q16 & Q17 & Q18 & Q19 & Q20 & Q21 & Q22 \\
\midrule
Students
& 0.836 & 0.531 & 0.603 & 0.699 & 0.921 & 0.918 & 0.968 & 0.968 & 0.924 & 0.944
& 0.740 & 0.563 & 0.820 & 0.827 & 0.783 & 0.429 & 0.484 & 0.487 & 0.374 & 0.552
& 0.672 & 0.658 \\
Chatbots
& 0.979 & 0.792 & 0.917 & 0.781 & 0.802 & 0.479 & 0.667 & 0.990 & 1.000 & 0.979
& 0.510 & 0.865 & 0.979 & 0.458 & 0.365 & 0.917 & 0.979 & 0.688 & 0.802 & 1.000
& 0.604 & 0.240 \\
\bottomrule
\end{tabular*}
\end{table*}

\subsection{Qualitative Analysis by Domain Experts}
\label{sec:method-RQ2} 
The qualitative analysis of the DIF items  was applied to the chemistry instrument. It was conducted by two subject-matter experts who lead a team responsible for analyzing the results of the high-stakes national matriculation exams in chemistry annually. 
The experts received a table listing the DIF properties of the items flagged by the LR-DIF method (e.g., POS or NEG; see \autoref{tab:lr_dif_summary}) and the instrument. 
The two experts independently reviewed the DIF items, integrated their observations through discussion, and provided qualitative judgments regarding their characteristics.

%% file: sections/experiments_and_results.tex
\section{Experiments and Results}\label{sec:results}
This section applies the pipeline outlined in Section~\ref{sec:method} and illustrated in Fig.~\ref{fig:pipeline}. The computational part was applied to the psychometric and chemistry data. 
The qualitative part was applied to the chemistry dataset, and is described in Subsection~\ref{sec:sme}.  We then conclude with the resulted procedure for detecting human:GenAI DIF items.

\subsection{Preprocessing.}\label{sec:preprocessing}
The following properties were computed \emph{per item $j$}. Let $X_{ij}\in\{0,1\}$ denote correctness for respondent $i$ on item $j$.

 \textbf{(i) Ability proxy.} We defined an \emph{ability proxy} for respondent $i$ as the \emph{rest score} $R_{i,-j}$, the respondent’s sum of correct answers across all items \textit{except} item $j$. Using rest score for respondent $i$, as opposed to using the global score of respondent $i$, avoids contamination from the target item and provides a consistent matching variable (i.e., a covariate used to condition comparisons on respondents’ overall proficiency).

  \textbf{(ii) Trimming non-overlapping ability ranges.} Following DIF analysis guidelines to ensure that both groups were compared only within the overlapping strata of the ability proxy, we removed the human responses in the ability ranges not populated by the chatbots. 

\textbf{(iii) ITC.} We computed the items' ITC values, which serve as indicators of item quality, and are later used as a diagnostic filter in the validation stage (see Subsection~\ref{sec:results-psych-diagnostics}). ITC values were computed on the human-only, pre-trimmed responses, as restricting to human data ensures item quality is evaluated relative to the reference group, while using pre-trimmed data preserves the full range of human ability for a fair discrimination estimate.

\subsection{DIF Analysis and Results}
We applied MH-DIF and LR-DIF on the data, as defined in Subsection~\ref{sec:dif-MH}, with humans and chatbots as reference and focal groups, respectively.

\textbf{1. MH-DIF.}
   For each item, we: \textbf{(a)} Stratified the trimmed responses by rest score into up to four quantile bins (or fewer when necessary). \textbf{(b)} Applied \texttt{difMH} to compute the MH statistics retrieved below. \textbf{(c)} Retrieved the common odds ratio $\alpha_{\text{MH}}$, the chi-square statistic, and its $p$-value. \textbf{(d)} Flagged items as DIF if $p<.05$ and the effect size category (defined in Section~\ref{sec:dif-MH}) was B or C. \textbf{(e)} Assigned direction: $\alpha_{\text{MH}}<1 \Rightarrow$ POS (chatbot advantage), $\alpha_{\text{MH}}>1 \Rightarrow$ NEG (chatbot disadvantage).
  
  \textbf{2. LR-DIF.}
  For each item, we: \textbf{(a)} Fit three nested LR models: (i)  Ability-only model. (ii) Uniform DIF model: add group membership as a predictor (iii)  Non-uniform DIF model: add the ability$\times$group interaction. \textbf{(b)} Computed likelihood-ratio test $p$-values for uniform ($p_{\text{uniform}}$) and non-uniform ($p_{\text{nonuniform}}$) DIF. \textbf{(c)} Calculated McFadden’s $\Delta R^2$ (fit improvement of the full model over the ability-only baseline). \textbf{(d)} Flagged items as DIF if $(p_{\text{uniform}}<.05$ or $p_{\text{nonuniform}}<.05)$ \emph{and} $\Delta R^2 > 0.035$. \textbf{(e)} Summarized flagged items by: (i) Log-odds contrast at median ability. (ii) Odds ratio $= \exp\{\text{log-odds contrast}\}$. (iii) Probability gap between humans and chatbots at median ability. \textbf{(f)} Classified direction as POS, NEG, or \emph{sign-change} if the group effect changed sign at different ability levels.

\input{sections/updated_MH_results_table}
\input{sections/updated_LR_results_table}
The items flagged as DIF by both methods are presented in Table~\ref{tab:mh_dif_summary} and Table~\ref{tab:lr_dif_summary}. As can be seen, MH-DIF flagged a large set of items as DIF, and particularly, a large number of items as POS (chatbot advantages conditioned on ability), including items on which the students scored higher when compared only on raw averages (e.g., items 6-7 in the chem. instrument; see average group scores in Table~\ref{tab:avg_score}). The LR-DIF was much more conservative, as expected, flagging less items overall. Its more sophisticated modeling enabled it also to identify items with non-uniform DIF (DIF magnitude changes across the ability ranges; e.g., chem. item 12).

These patterns indicate that chatbot–human differences are not only widespread but also vary by item, by direction, and by ability levels, highlighting the diagnostic power offered by DIF analysis. However, in light of the considerable disagreement between the methods (e.g., chemistry item 1, which was flagged as POS by the LR-DIF and NEG by the MH-DIF), we conducted Negative Control Analysis to discriminate true DIF from statistical artifacts.  

\input{sections/fp_counts_table}

\subsection{Negative Control Analysis}\label{sec:NCA-results}
For both datasets, we created $R=50$ \emph{null datasets} in which the chatbot group (n=60) was replaced by a random sample of human respondents (without replacement). We then reran the MH-DIF and LR-DIF pipelines on each null dataset. 
\autoref{tab:null-fp-matrix} reports the false-positive counts of items that were flagged at least once under the 50 placebo test runs. Two patterns stand out: \textbf{(1)} MH-DIF exhibits diffuse background noise: many items are flagged once or twice, and several reach 8–10\% false-positive rates. This noise profile appears on both instruments, indicating that MH-DIF alone cannot yield reliable DIF signals without additional filtering. \textbf{(2)} LR-DIF is far more stable. On the psychometric instrument it produced no null flags across all 50 runs, demonstrating excellent specificity. On the chemistry instrument, LR-DIF was generally stable but showed a small cluster of unstable items (e.g., Item 5).

These findings support LR-DIF as the primary DIF detection method. Yet the presence of a small unstable chemistry items cluster, which showed elevated false-positive rates in the null runs, needs additional item-level analysis and filtering (see Section \ref{sec:results-psych-diagnostics}).

\subsection{Psychometric Diagnostics and Results}\label{sec:results-psych-diagnostics}
We incorporated psychometric diagnostics to analyze and refine the candidate DIF set flagged by LR-DIF. As shown in \autoref{tab:null-fp-matrix}, Negative Control Analysis confirmed LR-DIF's stability, though chemistry items 5,9, and 10 exhibited false-positive rates above 10\%. To investigate these items further, we examined their ITC, and found that each had very low ITC values (Item 5 = 0.11, Item 9 = 0.08, Item 10 = 0.04). In contrast, chemistry items with ITC values above 0.20 were consistently stable across null runs. 
These results reinforced ITC as a reliable diagnostic for DIF analysis, using the standard 0.20 as a cutoff for `acceptable' \cite{shete2015item}.  
\subsection{RQ2: Domain Expert Analysis}\label{sec:sme}
As described in Subsection~\ref{sec:method-RQ2}, the qualitative analysis of the DIF items was conducted by two subject-matter experts who received the instrument and a table that listed the properties of the LR-DIF items, taken from Table~\ref{tab:lr_dif_summary}.

As shown in the Table \ref{tab:lr_dif_summary}, Items 12, 16, 17, and 19 had POS DIF, meaning that chatbots had \textit{higher} likelihood of answering them correctly than humans at the same ability level. In 12, the system successfully circumvented a common alternative conception according to which sodium chloride is composed of atoms. Instead, it correctly distinguished between the ionic lattice structure of NaCl and the behavior of its constituent particles in solution, emphasizing that dissolution involves the separation of Na$^{+}$ and Cl$^{-}$ ions rather than the breaking of atoms or the formation of atomic species. In 16, which involves a sequence of solution manipulations, the system effectively maintained consistency by monitoring the relationship between solute moles, solution volume, and concentration across several experimental stages. In 17 demonstrated a scientifically appropriate interpretation of the law of conservation of matter by situating the assessment of mass change at the level of the entire experimental system, rather than restricting the analysis just to the original metal. Item 19 further highlighted the system’s capacity to apply algorithmic reasoning: the multiple-choice structure facilitated accurate assignment of oxidation states, identification of electron transfer, and correct designation of oxidizing and reducing agents. Collectively, these examples underscore the system’s strength in tasks that prioritize rule-governed reasoning and conceptual clarity, where adherence to canonical chemical principles prevents susceptibility to intuitive but scientifically inaccurate interpretations.

In contrast, items 6, 7, 14, and 15, had NEG DIF, meaning that chatbots had \textit{lower} likelihood of answering them correctly than humans at the same ability level. These items place greater demands on visual interpretation, linguistic nuance, and the execution of complex, multi-step procedures. Items 6-7 require learners to interpret partial microscopic models, such as diagrams omitting solvent molecules, or to map symbolic labels (A–C) onto microscopic and macroscopic representations of aqueous solutions. These forms of representation, which depend heavily on implicit visual conventions, posed substantial challenges for the system, whereas students were able to draw upon prior representational. 14 illustrates a further point of divergence: the solution involves a multi-stage stoichiometric calculation, proceeding from gas volume to moles, through mole ratios, and finally to mass determination, with each step contingent on the correctness of the one before it. Such extended procedures are particularly vulnerable to intermediate errors in algorithmic reasoning. 15 compounds this difficulty, as errors originating in the previous calculation can propagate forward, whereas students often display the metacognitive awareness required to detect and correct inconsistencies in their own work. 
Taken together, these findings suggest that the chatbots excel in domains governed by formalized conceptual rules. By contrast, students retain a relative advantage in tasks that demand flexible interpretation of visual representations, sensitivity to linguistic subtleties, and careful monitoring of extended quantitative reasoning processes.

\noindent\textbf{\textit{Resulting Method.}}
The conclusion from the experimental results, which both validated the computational process statistically, and evaluated the meaningfulness of the LR-DIF results with subject matter experts on a sample dataset (chemistry), yielded that \texttt{applying LR-DIF to items with ITC $\ge$ 0.2 is a reliable method for identifying items that, with high confidence, exhibit differential behavior across humans and chatbots}.

%% file: sections/updated_MH_results_table.tex

\begin{table}[!t]
\centering
\caption{\texorpdfstring{MH-DIF: Summary of DIF-flagged items.}{MH-DIF: Summary of DIF-flagged items.}}
\begin{threeparttable}

\label{tab:mh_dif_summary}

\footnotesize
\setlength{\tabcolsep}{4pt}
\renewcommand{\arraystretch}{1.15}

\begin{tabularx}{\linewidth}{lXX}
\toprule
Inst. & POS & NEG \\
\midrule
\textbf{Psych.} &
4-6, 9-10, 12, 16, 21-22, 25, 28, 30-33 &
11, 13-15, 17, 20, 23-24, 34-38, 40 \\
\textbf{Chem.} &
5-7, 11, 14-15, 21-22 &
1-3, 12-13, 16-19 \\
\bottomrule

\end{tabularx}
\end{threeparttable}
\end{table}

%% file: sections/updated_LR_results_table.tex
\newcommand{\uI}[1]{\ensuremath{\bar{#1}}}
\newcommand{\nI}[1]{\ensuremath{\tilde{#1}}}

\begin{table}[!t]
\centering
\caption{LR-DIF flagged items: $\uI{Q}$: uniform; $\nI{Q}$: non-uniform.}
\begin{threeparttable}

\label{tab:lr_dif_summary}

\footnotesize
\setlength{\tabcolsep}{4pt}
\renewcommand{\arraystretch}{1.15}

\begin{tabularx}{\linewidth}{lYY}
\toprule
Inst. & POS & NEG  \\
\midrule
\textbf{Psych.} &
\uI{11}, \nI{13}-\nI{15}, \nI{34}, \uI{40} &
\nI{16}, \uI{21}, \nI{25}, \nI{31} \\
\textbf{Chem.} &
\nI{12}, \uI{16}, \uI{17}, \uI{19} &
\uI{5}-\uI{7}, \nI{14}, \uI{15}, \nI{22} \\
\bottomrule
\end{tabularx}

\end{threeparttable}
\end{table}

%% file: sections/fp_counts_table.tex

\begin{table*}[!t]
\centering
\caption{Negative Control Analysis: The items on each \textit{false\_positive\_count} level (0 omitted).}
\label{tab:null-fp-matrix}

\scriptsize
\renewcommand{\arraystretch}{1.00}
\setlength{\tabcolsep}{2pt}
\begingroup
\linespread{0.95}\selectfont

\begin{tabular*}{\textwidth}{@{\extracolsep{\fill}}%
>{\raggedright\arraybackslash}p{0.06\textwidth}%
>{\raggedright\arraybackslash}p{0.42\textwidth}%
>{\raggedright\arraybackslash}p{0.24\textwidth}%
|%
>{\raggedright\arraybackslash}p{0.12\textwidth}%
>{\raggedright\arraybackslash}p{0.12\textwidth}%
@{}}
\toprule
\textbf{Counts}
& \textbf{Psych.(MH-DIF)} & \textbf{Chem.(MH-DIF)}
& \textbf{Psych.(LR-DIF)} & \textbf{Chem. (LR-DIF)} \\
\midrule

1  & 6-7, 10, 12-15, 17, 19-21, 24, 27-28, 33, 38
   & 2-3, 6, 9, 13-14, 17-19, 22
   & -- & 14 \\

2  & 9, 18, 22, 29-32, 37, 39
   & 7, 16
   & -- & 6-8 \\

3  & 2, 8, 16, 26, 34
   & 11, 20, 21
   & -- & -- \\

4  & 25, 35
   & -- &   -- & -- \\

5  & 23, 40
   & 12 &   -- & -- \\

6-7  & -- & -- &   -- & -- \\

8-11  & -- &   -- & -- & 5, 9-10 \\

\bottomrule
\end{tabular*}
\endgroup
\end{table*}

%% file: refs.bib
@String{Academic = "Academic Press" }

@String{Springer = "Springer-Verlag" }

@inproceedings{liu2025llms,
  title={Do LLMs Make Mistakes Like Students? Exploring Natural Alignments Between Language Models and Human Error Patterns},
  author={Liu, Naiming and Sonkar, Shashank and Baraniuk, Richard},
  booktitle={International Conference on Artificial Intelligence in Education},
  pages={364--377},
  year={2025},
  organization={Springer}
}

@article{yan2024practical,
  title={Practical and ethical challenges of large language models in education: A systematic scoping review},
  author={Yan, Lixiang and Sha, Lele and Zhao, Linxuan and Li, Yuheng and Martinez-Maldonado, Roberto and Chen, Guanliang and Li, Xinyu and Jin, Yueqiao and Ga{\v{s}}evi{\'c}, Dragan},
  journal={Br. J. Educ. Technol.},
  volume={55},
  number={1},
  pages={90--112},
  year={2024},
  publisher={Wiley Online Library}
}

@article{rogers1993comparison,
  title={A comparison of logistic regression and Mantel-Haenszel procedures for detecting differential item functioning},
  author={Rogers, H Jane and Swaminathan, Hariharan},
  journal={Applied psychological measurement},
  volume={17},
  number={2},
  pages={105--116},
  year={1993},
  publisher={Sage Publications Sage CA: Thousand Oaks, CA}
}

@article{eggers2024placebo,
  title={Placebo tests for causal inference},
  author={Eggers, Andrew C and Tu{\~n}{\'o}n, Guadalupe and Dafoe, Allan},
  journal={Am. J. Polit. Sci.},
  volume={68},
  number={3},
  pages={1106--1121},
  year={2024},
  publisher={Wiley Online Library}
}

@article{shete2015item,
  title={Item analysis: An evaluation of multiple choice questions in Physiology examination},
  author={Shete, Anjali},
  journal={J. of Contemporary Med. Educ.},
  year={2015}
}

@inproceedings{echterhoff-etal-2024-cognitive,
    title = "Cognitive Bias in Decision-Making with {LLM}s",
    author = "Echterhoff, Jessica Maria  and
      Liu, Yao  and
      Alessa, Abeer  and
      McAuley, Julian  and
      He, Zexue",
    booktitle = "Findings of the Association for Computational Linguistics: EMNLP 2024",
    year = "2024",
    publisher = "Association for Computational Linguistics",
}

@misc{yacobson2025benchmarking,
  author       = {Yacobson, Elad and Schleifer, Yael and Bar-Dov, Ziva and Rap, Shelley and Blonder, Ron and Alexandron, Giora},
  title        = {Benchmarking LLMs vs. High School Students on Standard Chemistry Exams: Insights for Science Education},
  year         = {2025},
  howpublished = {OSF Preprints},
  url          = {https://doi.org/10.35542/osf.io/3n7xr_v1}
}

@inproceedings{khalil2023will,
  title={Will {ChatGPT} Get You Caught? {Rethinking} of Plagiarism Detection},
  author={Khalil, Mohammad and Er, Erkan},
  booktitle={International Conference on Human-Computer Interaction},
  pages={475--487},
  year={2023},
  organization={Springer}
}

@inproceedings{Strugatski_25,
  author       = {Alona Strugatski and
                  Giora Alexandron},
  title        = {Applying {IRT} to Distinguish Between Human and Generative {AI} Responses
                  to Multiple-Choice Assessments},
  booktitle    = {Proceedings of the 15th International Learning Analytics and Knowledge
                  Conference, {LAK} 2025,  2025},
  pages        = {817--823},
  publisher    = {{ACM}},
  year         = {2025},
}

@online{The-AI-Cheating-Crisis,
  author = {Akbari, Noor},
  title = {The {AI} cheating crisis: {Education} needs its anti-doping movement},
  year = {2024},
  note = {Retrieved from \url{https://www.edweek.org/technology/opinion-the-ai-cheating-crisis-education-needs-its-anti-doping-movement/2024/02}},
}

@online{Prothero24,
  author = {Arianna Prothero},
  title = {New Data Reveal How Many Students Are Using {AI} to Cheat},
  year = {2024},
  url={https://www.edweek.org/technology/new-data-reveal-how-many-students-are-using-ai-to-cheat/2024/04}
}

@Article{susnjak2024chatgpt,
AUTHOR = {Susnjak, Teo and McIntosh, Timothy R.},
TITLE = {{ChatGPT}: {The} End of Online Exam Integrity?},
JOURNAL = {Education Sciences},
VOLUME = {14},
YEAR = {2024},
NUMBER = {6},
ARTICLE-NUMBER = {656}
}

@article{sorenson2024identifying,
  title={Identifying Generative Artificial Intelligence Chatbot Use on Multiple-Choice, General Chemistry Exams Using {Rasch} Analysis},
  author={Sorenson, Benjamin and Hanson, Kenneth},
  journal={Journal of Chemical Education},
  volume={101},
  number={8},
  pages={3216--3223},
  year={2024},
  publisher={ACS Publications}
}

@article{borges2024could,
  title={Could ChatGPT get an engineering degree? Evaluating higher education vulnerability to AI assistants},
  author={Borges, Beatriz and and others},
  journal={Proceedings of the National Academy of Sciences},
  volume={121},
  number={49},
  year={2024},
  publisher={National Academy of Sciences},
}

@inproceedings{wang2024examining,
  title={Examining the potential and pitfalls of {ChatGPT} in science and engineering problem-solving},
  author={Wang, Karen D and Burkholder, Eric and Wieman, Carl and Salehi, Shima and Haber, Nick},
  booktitle={Frontiers in Education},
  volume={8},
  pages={1330486},
  year={2024},
  organization={Frontiers Media SA}
}

@book{DeAyalaIRT,
  title={The theory and practice of item response theory},
  author={De Ayala, Rafael Jaime},
  year={2013},
  publisher={Guilford Publications}
}

@article{Cotton03032024,
author = {Debby R. E. Cotton and Peter A. Cotton and J. Reuben Shipway and},
title = {Chatting and cheating: Ensuring academic integrity in the era of {ChatGPT}},
journal = {Innovations in Education and Teaching International},
volume = {61},
number = {2},
pages = {228--239},
year = {2024},
publisher = {SRHE Website},
doi = {10.1080/14703297.2023.2190148}
}

@article{lu2022learn,
  title={{Learn to Explain: Multimodal Reasoning via Thought Chains for Science Question Answering}},
  author={Lu, Pan and others},
  journal={Advances in Neural Information Processing Systems},
  volume={35},
  pages={2507--2521},
  year={2022}
}

@article{wang2023scibench,
  title={{Scibench: Evaluating College-Level Scientific Problem-Solving Abilities of Large Language Models}},
  author={Wang, Xiaoxuan and Hu, Ziniu and Lu, Pan and Zhu, Yanqiao and Zhang, Jieyu and Subramaniam, Satyen and Loomba, Arjun R and Zhang, Shichang and Sun, Yizhou and Wang, Wei},
  journal={arXiv preprint arXiv:2307.10635},
  year={2023}
}

@article{Martinkov2017DIF,
author = {Martinkov\'{a}, Patr\'{\i}cia and Drabinov\'{a}, Ad\'{e}la and Liaw, Yuan-Ling and Sanders, Elizabeth A. and McFarland, Jenny L. and Price, Rebecca M.},
title = {Checking Equity: Why Differential Item Functioning Analysis Should Be a Routine Part of Developing Conceptual Assessments},
journal = {CBE Life Sci. Educ.},
volume = {16},
number = {2},
year = {2017}
}

@article{mantel1959statistical,
  title={Statistical aspects of the analysis of data from retrospective studies of disease},
  author={Mantel, Nathan and Haenszel, William},
  journal={Journal of the national cancer institute},
  volume={22},
  number={4},
  pages={719--748},
  year={1959},
  publisher={Oxford University Press}
}

@article{magis2010general,
  title={A general framework and an R package for the detection of dichotomous differential item functioning},
  author={Magis, David and Beland, Sebastien and Tuerlinckx, Francis and De Boeck, Paul},
  journal={Behavior research methods},
  volume={42},
  number={3},
  pages={847--862},
  year={2010},
  publisher={Springer}
}

@online{zwick2012rr1208,
  author      = {Zwick, Rebecca},
  title       = {A Review of ETS Differential Item Functioning Assessment Procedures: Flagging Rules, Minimum Sample Size Requirements, and Criterion Refinement},
  institution = {Educational Testing Service},
  address     = {Princeton, NJ},
  number      = {RR-12-08},
  type        = {ETS Research Report},
  year        = {2012},
  doi         = {10.1002/j.2333-8504.2012.tb02290.x},
  url         = {https://files.eric.ed.gov/fulltext/EJ1109842.pdf}
}

@article{JodoinGierl2001,
  author    = {Jodoin, Michael G. and Gierl, Mark J.},
  title     = {Evaluating Type I Error and Power Rates Using an Effect Size Measure with the Logistic Regression Procedure for {DIF} Detection},
  journal   = {Applied Measurement in Education},
  year      = {2001},
  volume    = {14},
  number    = {4},
  pages     = {329--349},
  doi       = {10.1207/S15324818AME1404_2},
  url       = {https://doi.org/10.1207/S15324818AME1404_2}
}

@article{koedinger2012automated,
  title={Automated Student Model Improvement.},
  author={Koedinger, Kenneth R and McLaughlin, Elizabeth A and Stamper, John C},
  journal={International Educational Data Mining Society},
  year={2012},
  publisher={ERIC}
}

@inproceedings{barnes2005q,
  title={The q-matrix method: Mining student response data for knowledge},
  author={Barnes, Tiffany},
  booktitle={American association for artificial intelligence 2005 educational data mining workshop},
  pages={1--8},
  year={2005},
  organization={AAAI Press, Pittsburgh, PA, USA}
}
